\documentclass[graybox]{svmult}

\usepackage{type1cm}        
\usepackage{makeidx}         
\usepackage{graphicx}        
\usepackage{multicol}        
\usepackage[bottom]{footmisc}

\usepackage{newtxtext}       %
\usepackage[varvw]{newtxmath}       

\usepackage{cite}
\usepackage{float}
\usepackage{scrextend} 
\usepackage{subcaption}
\usepackage{url}
\usepackage{verbatim}           
\usepackage{xcolor}
\usepackage{xspace}

\usepackage[pagebackref=true,colorlinks=true]{hyperref}
\hypersetup{
     unicode=true,          
     pdfnewwindow=true,      
     bookmarksnumbered=true,
     colorlinks=true,       
     linkcolor=blue,          
     citecolor=blue,        
     filecolor=magenta,      
     urlcolor=cyan,           
     menucolor=blue,
     linktocpage=true
}

\def\TeV{\ifmmode {\mathrm{\ Te\kern -0.1em V}}\else
                   \textrm{Te\kern -0.1em V}\fi}%
\def\GeV{\ifmmode {\mathrm{\ Ge\kern -0.1em V}}\else
                   \textrm{Ge\kern -0.1em V}\fi}%

\makeindex             

\newcommand{\be}{\begin{equation}}
\newcommand{\ee}{\end{equation}}
\newcommand{\bea}{\begin{eqnarray}}
\newcommand{\eea}{\end{eqnarray}}
\newcommand{\ba}{\begin{eqnarray}}
\newcommand{\ea}{\end{eqnarray}}

\begin{document}
\title*{The quest for a Quark-Gluon Plasma}


\author{Edward Shuryak}

\institute{Edward Shuryak
\at Center for Nuclear Theory, Department of Physics and Astronomy, \\
Stony Brook University, Stony Brook, 
New York 11794–3800, USA; \\
\email{edward.shuryak@stonybrook.edu}
}

%
%
\maketitle



\abstract{
An extended summary of the field can be found in
large books and reviews, e.g., my take in Ref.~\cite{Shuryak:2024zrh}. It is hard
 to compress  its 600+ pages and 50+ years of research to a few given below.  Those include
only recollection of few episodes, from early days to now,  showing
how vague dreams  were eventually turning into solid and detailed scientific facts.
}


\section{Before Quark-Gluon Plasma (QGP)}
My road into physics went through 1970-1974  ``aspirantura" (graduate school), in Budker Institute of Nuclear Physics, Novosibirsk (Siberia). Let me mention a
few papers of this period related to high-energy collisions and hadronic matter.  In Ref.~\cite{Shuryak:1972bdm}, it was shown that 
particle momentum spectra 
in particular channels of $p\bar p\rightarrow mesons$ are well described by the statistical model by Pomeranchuck, based on freeze-out at 
fixed particle density. It introduced strangeness suppression following micro-canonical distribution, later rediscovered by Redlich. Further development was usage of ``improved thermodynamics", in which Sterling
formula for factorial the subleading $O(1/N)$ term was
retained.
Ref.~\cite{Shuryak:1972kq} was devoted to two-meson
interference effects (later called ``femtoscopy), and a general theory based on
a ``statistical source function" was developed.

With my first student, 
O.V. Zhirov~\cite{Zhirov:1975qu}, we
tried  to relate the experimental data on multiparticle
production to Landau's hydrodynamical model. First, following Landau, 
we focused on rapidity distribution, which surprisingly worked well. Then we were also
trying to find the {\em transverse flow} by comparing $\pi,K,p$ spectra. That was a disappointment: only
 a decade later, heavy ion collisions came into being and
displayed it. (Its hydrodynamical description
was worked out  two decades later with my first Stony Brook
student, C.M.~Hung~\cite{Hung:1994eq}.)

\section{Why did I call it ``quark-gluon plasma"?}
In the early 1970's theorists were already discussing possible 
``abnormal" states of matter.  One notable proposal, by Lee and Wick, suggested a transition to a phase in which the scalar field vacuum expectation value (VEV) vanishes, rendering nucleons - essentially constituent quarks — effectively massless.
Around the same time, pion and kaon condensates were also actively discussed at respective meetings\footnote{Needless
to say, I was not able to attend any of them.}.

Yet everything changed after the discovery of the {\em asymptotic freedom} by Gross, Wilczek and Politzer in 1973.
It became obvious that at very high $T$ or $\mu$ (temperatures or chemical potentials) matter
must be  in a {\em weak coupling regime}, in which 
 all nonperturbative phenomena -- confinement and chiral symmetry breaking in particular --
cannot possibly occur. Thus, in such limit the QCD matter must return to 
its ``normal" state, made of quarks and gluons, without condensates. 

 By that time, it was known ``in principle" how to evaluate perturbative Feynman diagrams at finite $T$ or $\mu$, and many people (me included) started to calculate perturbative
 corrections to energy/pressure. It was rather simple to calculate the
first few diagrams for gluon-gluon or quark-gluon scattering. Yet that could not really be continued without
 addressing  the main issues,  the so-called ``infrared completion" of the theory. Recall that "asymptotic freedom" is an $anti$-screening of a charge at small distances, by
virtual gluons in the vacuum, opposite to $screening$ in QED. 
What long-distance behavior would occur in hot QCD matter?

 There remained several
  nontrivial issues here to be resolved. The first (technical but far from simple) 
  was the choice of the gauge: one can either follow Gross et al and 
  use covariant gauges with Faddeev-Popov ghosts, or use 
  non-covariant (e.g., Coulomb) gauge
  in the matter rest frame. I opted for the latter, which was closer to physics but required the
  development of novel
   Feynman rules. In it, there are  novel
  propagators for electric and magnetic fields, with the corresponding polarization tensors 
    \begin{eqnarray}
  D_{00}={1 \over \vec{k}^2+\Pi_{00}(\omega,k,T)} \qquad D_{mn}=- {\delta_{mn}-k_m k_n/
 k^2  \over \omega^2 -\vec{k}^2-\Pi_\perp(\omega,k,T)},\,\,\, 
 \nonumber \\
 M_E^2=\Pi_{00}(\omega=0,k \ll T)=g^2 T^2 (1+N_f/6)
 \end{eqnarray}
These  propagators and  polarization tensors 
 are generally gauge dependent, in general to be combined with vertex renormalization
 in order to get physical running of the charge. 
 Fortunately, {\em in the Coulomb gauge}, one can show that
 there is $no$ vertex charge correction at small $k,\omega\ll T$, 
 so for example, the {\em electric screening mass} $M_E^2$ (given above) is gauge independent. My calculation
 has shown it to be $positive$, generating  $screening$ of
 the electric charge. Other limits of $\Pi_{00}$ also
 provided valuable insights: for example at $k=0, \omega \ll T$
I found that a gluon does develop a ``plasmon pole", same
as in ordinary electrodynamical plasmas. Returning from
Euclidean time to Minkowskian one, I  also found gluon
quasiparticles well equipped by a  ``Landau damping" also known in 
ordinary plasmas. So, in 1976 I sent a large paper \cite{Shuryak:1977ut} to JETP 
which announced that the {\em normal phase of hot QCD } must be a ``hadronic plasma"\footnote{ 
Let me object some recollections saying 
 ``Shuryak coined the name Quark-Gluon Plasma...". No, I was not playing with words here, but actually
 $derived$  several key features of this phase 
 common with electrodynamical plasmas. It is not a name but a statement.}. In  my next paper~\cite{Shuryak:1978ij}, its name changed to Quark-Gluon Plasma (QGP), which became the standard terminology ever since and used widely\footnote{When INSPIRE  introduced ``search by a word" in  its database it pick up ``QGP" as an example, revealing tens of thousands of papers containing it.}.

Let me briefly mention  part of the story related to another polarization operator describing the
{\textbf magnetic sector}. My (and subsequent)
perturbative calculations had shown that  magnetic screening mass does   $not$ appear, and therefore, power infrared divergences  
in magnetic sector remained uncured.  
Polyakov~\cite{Polyakov:1979gp}  has then suggested
    that  this  disease can
    still be
    cured by the development of some magnetic screening  length  for  the
    gluomagnetic field, of higher order of $M_M\sim g^2T$.
  If so, using it as a IR cutoff one would conclude
that {\em all diagrams} above the 8th order  contain a
contribution of the same magnitude
    $\delta\Omega\sim g^6 T^4 $.
Decades later, lattice studies have shown that Polyakov's magnetic mass does indeed exists.
Ref.~\cite{Linde:1980ts}  suggested that magnetic screening mechanism can be due to
  magnetic t'Hooft-Polyakov monopoles\footnote{ By then I never met Andrei Linde,  and he sent me a very polite letter asking basically if I checked my calculation well enough. My  answer, also polite, was
  in essence: "why don't you check it yourself?"}. Kajantie school   related magnetic sector  to an effective theory,
 in turn related to confining 3-dimensional Yang-Mills theory. 

Crucial  development in 1970's was Ken Wilson's lattice formulation of the nonabelian gauge theory,
eventually leading to numerical simulations of the gauge vacuum. Jumping over decades of hard work by many, it was found
that the critical temperature $T_c^{QCD}\approx 155$~MeV. Less widely known
are lattice results for values of electric and magnetic screening masses above $T_c$
$ M_E/T\approx 7\pm 0.5, \,\, M_m/T\approx 4.6\pm 0.5$.
While indeed positive, they are so large that their description by perturbative expressions
makes little sense\footnote{Even at electroweak phase transitions $T_c^{ew}\approx 1000 T_c^{QCD}$ one finds $g\approx 1$, not small.}. The same problem is for kinetic quantities, such as viscosities: the
perturbative results did not match the empirical values, obtained decades later. Eventually, it was recognized
that the QGP created in experiments is a ``strongly coupled" regime, sometimes called, ``sQCD".
Its theoretical description required novel tools  different from perturbative QCD: but this part
does not belong to this collection.

\section{Heavy ion collisions: potential observables}

In the second half of 1970's it became clear that, in order to get experimental ``quest for QGP" really going, one
needed to focus on specific observables. A sequence of my papers devoted to those started with a letter publication\footnote{``Psions" in its title are the charmonia states: discussion of their excitation in thermal medium  predated the famous 1986 Matsui-Satz work of charmonium ``unbinding".
} \cite{Shuryak:1978ij},
larger article and eventually a review in Physics Reports~\cite{Shuryak:1980tp}. It summarized theoretical status of the  ``QCD
at finite temperatures", but also took a much wider look at possible experimental program to be developed.

Suggestions included possible observations of ``penetrating probes" - photons or dileptons - and presented the 
first QCD calculations of their  emission rates.  Also production
of new flavors - strangeness and charm - from thermal gluons was evaluated\footnote{Again, predated influential 1982 strangeness paper by Muler and Rafelski.}. Collective flows and especially hydrodynamical approach to fireball  
explosion. It included discussion of rapidity-independent (``scaling") scenario, a precursor
to 1983 Bjorken's famous explicit solution. 

My ``confinement" in Siberia ended in 1982, when, unlike previous invitations, the invitation to CERN was suddenly approved. Arriving to CERN, I discovered
that in fact I was invited by 
``educational program" and was supposed to give three lectures in big Auditorium for hundreds of CERN staff members. Lectures have been written promptly, in a couple of days and also promptly appeared as ``CERN Yellow report". 
Quest for QGP was in the air, it was supposed to include conversion of the first hadronic collider ISR into heavy ion project,  a precursor to RHIC/LHC.
While unfortunately it did not happen, those discussions significantly strengthened  the community\footnote{I presume
this event was initiated by Prof. van Hove, CERN Director General, with whom I had several discussions. One day, a
distinguished-looking gentleman came to my office there,
 turned out to be Prof. Weisskopf, another former  CERN Director General. He asked what I am doing, and I enthusiastically start explaining my initial thoughts about
the instanton vacuum. ``This is too confusing" said Weisskopf, "but I met in the corridor my former student
and if you explain it to him he can probably explain it to me".
In a moment, he returned with another distinguished-looking gentleman, who introduced himself simply as "Murray Gell-Mann".}.

\section{The hydro story}
A ``big question" pending over all of this 
was whether the fireball produced in high-energy 
collisions would be made of ``matter", or just a ``firework"
of quarks and gluons. Specifically, the question is whether  high energy heavy-ion collisions will produce a ``macroscopic" object, large
compared to
 the ``mean free path",  $R\gg l_{mfp}$.  Most people did not
 believe it, and 
 used perturbative kinetic theory
(``partonic cascades") to disprove it\footnote{Attempts to derive QGP hydro from kinetics appear and disappear all the time.
It makes little sense, precisely because they constitute two opposite regimes. Attempts to justify Boltzmann zero correlation hypothesis, the basis for kinetics, do  fail repeatedly.}. Eventually, hydrodynamics triumphed, and the
extracted viscosity was found to be too small for perturbative theory.

 Results of early heavy ion experiments at CERN SPS and Brookhaven AGS in 1980-1990's were described by hadronic cascade
event generators (RQMD, UrQMD etc). Yet 
observed radial flow was also   described by
hydro \cite{Hung:1994eq} as well. At the crucial moment 
(the end of 1990's) with yet another Stony Brook student we made hydro predictions for future RHIC runs. The title of the paper in Ref.~\cite{Teaney:2000cw} was: "Flow at the SPS and RHIC as a quark gluon plasma signature". Basically, while cascades predicted elliptic flow to be a factor 2 weaker at RHIC relative to SPS, our hydro predicted the opposite, that $v_2$
parameter will  be twice larger! The first days of RHIC run have 
confirmed the hydro predictions\footnote{Similar story repeated a decade later, before the first heavy ion run of LHC:
we argued that elliptic (and higher) flows will increase further, while pessimists once again suggested the opposite.}.
Experimental data on elliptic flow not only confirmed our hydro predictions,  but in fact, the $v_2$ parameter was linearly growing  with transverse momentum
 well beyond the region  $p_\perp < 1$~GeV in which we dared to
  calculate. Its growth continues for several orders of magnitude down the spectra, 
 till  at $p_\perp \approx 3-4$~GeV, at which ellipticity
 reaches a magnitude so large that it was seen by eye, on the event-by-event basis! 
 
Pions with such $p_\perp$ have quite significant transverse rapidity,  which can only be generated
in a narrow ``outer rim" of the fireball. This observation
implies by itself that the mean free path must be small enough. And indeed,
it turned out impossible 
to reproduce this ``hydro splash" effect by any ``parton cascades", even with 
unrealistically huge  cross sections \cite{Molnar:2001nk}. 

Let me briefly mention further hydro triumphs.
Following Ref.~\cite{Teaney:2003kp}, the magnitude of
shear viscosity $\eta$ can be inferred from
elliptic flow at $p_\perp \approx 3-4$~GeV, and it 
 turned out to be surprisingly small. The precise value was hard to get unless one had 
 theory predictions~\cite{Staig:2010pn} and measurements of higher harmonics.
Suggested "acoustic systematics" formula for higher harmonics
\be
{v_n \over \epsilon_n}
 = {\rm exp}\left(-C  {\eta \over s} {n^2  \over T \bar{R}}  \right)
 \label{eqn_visc_filter} \ee
($C,\eta,s,n,T,\bar{R}$ are a constant, shear viscosity, entropy density, harmonic number and mean fireball size, respectively.)
 described a very wide range of data \cite{Lacey:2013is}. The obtained  viscosity-to-entropy ratio have shown that the QGP  is a ``near-perfect liquid",  not a gas.

Furthermore, motivated by maxima and minima in spectral function of cosmological microwave background, we calculated 
phases of higher and higher harmonics, and
predicted that the phase factor will lead to a minimum at the 7-th harmonics (left plot from Ref.~\cite{staig:2011wj}). People
were laughing at this prediction, and yet, a decade later,  ALICE  (right plot)~\cite{ALICE:2020sup} confirmed it. So, not only explosion is hydrodynamical, but perturbations on top of it are in fact sound waves.

\begin{figure}[h!]
  \begin{center}
  \includegraphics[width=0.43\textwidth]{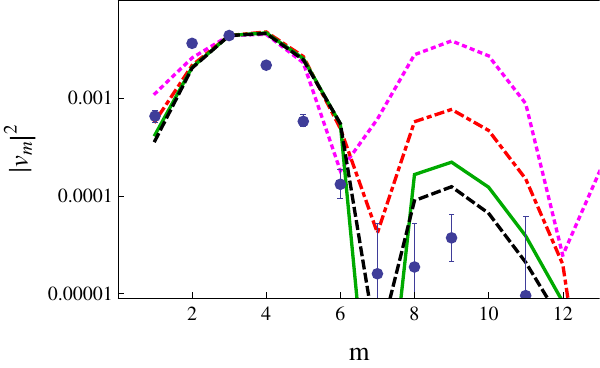}
  	\includegraphics[width=0.5\textwidth]{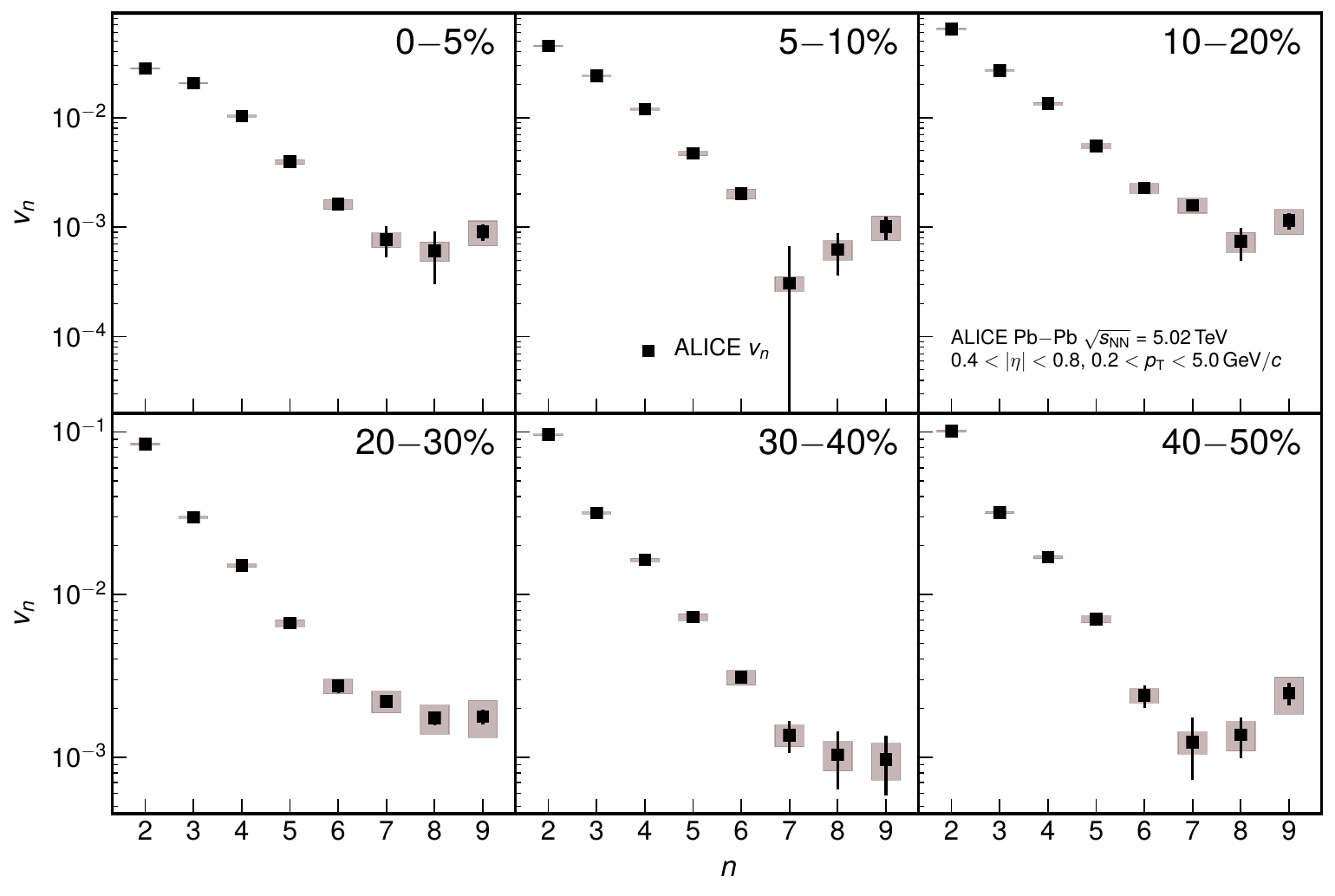}
   \caption{
   {{\it (Left)}}: the mean squared amplitude of harmonics $\langle v_m^2 \rangle$ versus its number $m$. The
   lines are from analytic linearized
   hydrodynamic calculations of the correlation function harmonics, $v_m^2$, based on 
   a Green function from a point source  for four values of viscosity  $4\pi\eta/s$ = 0,1,1.68,2 (top to bottom)~\cite{staig:2011wj}.
   {{\it (Right)}}: Experimental data on higher harmonics ($v_n$) as a function of harmonic order $n$ for various centrality intervals~\cite{ALICE:2020sup} measured by ALICE.
   }
  \label{fig_vn_comp}
  \end{center}
\end{figure}

Furthermore, not only ``large"
fireballs (created by Pb-Pb or Au-Au collisions), but also  ``small systems" 
in  $pA$ (and even in high multiplicity $pp$ collisions)  show well-developed radial, elliptic and triangular
flows as well. RHIC data \cite{PHENIX:2022nht} for d-Au and 
$^3$He-Au
were subsequently well described by hydro calculations.  

\section{Dileptons and photons from QGP }
While secondary hadrons come from ``freeze-out surfaces",
the ``penetrating probes" (photons and dileptons) come from
integrated emission rates over the whole 4-d volume\footnote{ Original suggestion \cite{Feinberg:1976ua} incorrectly used black-body expressions for them, but correctly emphasized enhancement proportional to fireball lifetime.}
The QGP paper \cite{Shuryak:1978ij} already included such rates, and emphasise that strong corrections to them must be small and thus the  ``penetrating probes" would provide a quantitative check on matter evolution.

Dileptons with masses $M_{l^+l^-}<1$~GeV show vector meson spectral densities,
 thus telling us if they melt in matter or not. The
 ``intermediate mass dileptons" $1<M_{l^+l^-}<3$~GeV observed by NA45, Helios-3 and NA38 at CERN  ascribed them initially to charm decays.
Yet in Ref.~\cite{Rapp:1999zw} we have shown that data are consistent with thermal dileptons from QGP emission. It took a heroic efforts of  the NA60 experiment (which put silicon tracking under full beam, ``killing it softly"!)
to prove  that they  did not come from charm  but from  QGP radiation.

Unlike photons, dileptons have both $p_\perp$ spectra (affected by flow) and invariant mass spectra (not affected), so they
also tell us about the timing of hydro flow development. Latest study focused
also on dilepton elliptic flow: to my knowledge here some puzzles remain.

\section{Deconfinement and chiral phase transitions}
Let me now jump from phenomenology to theory. A mainstream numerical simulations on the lattice gradually progressed,  from pioneering papers in 1980's discovering deconfinement in gauge theory, to quantitative control of
QCD thermodynamics. Higher order  susceptibilities are calculated and related to event-by-event
fluctuations, as was suggested in Ref.~\cite{Shuryak:1997yj}. Yet
 we would like to formulate the physics of these phase transitions understandable to a wide range of physicists,
 in high-energy and condense matter community.
 
 One pillar of such explanation is electric-magnetic duality. Since the time of Maxwell it was noticed
 by many that electric and magnetic sectors show
 puzzling asymmetry: there are no magnetic charges and currents. Famous Dirac's effort to reconcile 
 magnetic monopoles with quantum mechanics brought
 Dirac's quantization of electric and magnetic charges
 $e$ and $g$: $$ e*g={n \over 2}, \,\,\, n\in integers $$
or otherwise Dirac strings would become visible.

All attempts to find QED monopoles so far failed, yet in
QCD-like theories a magnetic monopole-like localized
object were located on the lattice. It was demonstrated
that they rotate around electric flux tubes, like
Cooper pairs do around magnetic flux tubes (known
as Abrikosov's vortices) in the superconductors. 
It was also demonstrated~\cite{D'Alessandro:2010xg}
 that lattice monopole undergo
   Bose-Einstein condensation, exactly at the deconfinement transition temperature.

The QCD coupling  $e$ ``runs" \footnote{Usually called  $g$ but not in this
section in which $g$ is the magnetic coupling.}, means it varies with the energy scale. In particular, at high $T$ $e(T)$ is small
and the phase is a  weakly coupled QGP, but it grows as the temperature is decreasing toward the deconfinement
transition. The Dirac condition then $requires$ that 
magnetic coupling $g(T)$ must decrease, with their product being constant\footnote{Demonstrated for lattice monopoles to happen. See Ref~\cite{Liao:2008jg}.}. All indicated that transition, 
QGP-hadronic gas, can
 be understood as a fight between electric (quarks and gluons) and magnetic (monopoles) degrees of freedom\footnote{Similar fight has been beautifully demonstrated in  supersymmetric theories,
 in which eventually one has a phase of weakly coupled monopoles, see Ref.~\cite{Seiberg:1994rs}.}!
Yet there remained one serious problem: semiclassical monopole solutions exist in other theories (e.g.Seiberg-Witten theory) with colored scalars\footnote{Known as 't Hooft-Polyakov monopoles.}, but not 
in QCD!  

Here comes the central idea: if one does not have 
4-d Lorentz scalars, can a 3d scalar defined in the
rest frame of matter be used instead? There is such
scalar, in fact the {\em order parameter of
deconfinement} known as the Polyakov loop,  
 $\hat P=\langle pexp[ig\int dx_0A_0] \rangle $.
Using it, P.~van~Baal and collaborators  found monopole-like solutions now known as {instanton-dyons}.
Unfortunately, these  monopoles use ``Matsubara  time" and Euclidean component of the gauge potential $A_0$, so they are not technically particles which one can put in a pocket. Their name shows they are descendants
of another topological beast, the instantons, or the tunneling events.

 After about the year 2000, it became clear\footnote{Unfortunately Pierre van Baal got a stroke and got incapacitated. In our last meeting he suggested me to work on it, but we did not get far as  the second stroke killed him. For few years the leader in this direction was Dmitry Diakonov, but he also suddenly died in 2012.} that in terms of
 instanton-dyons one perhaps would be able to describe
 both the deconfinement and the chiral phase transitions in QCD. And indeed
(not much noticed by the mainstream), a quite successful {\em semiclassical theory of both transitions} were developed around 2020,  reproducing all lattice 
phase transitions, without and with dynamical quarks. 

\begin{figure}[htbp]
\begin{center}   
\includegraphics[width=0.4\textwidth]{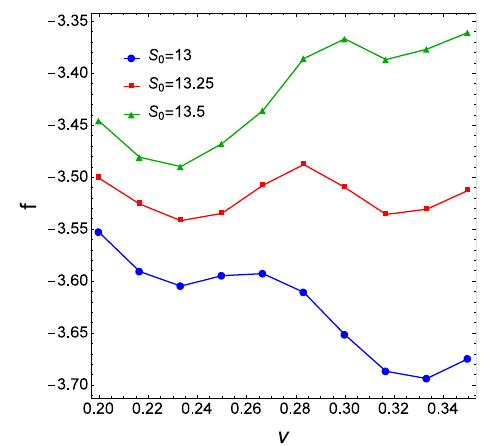}
\includegraphics[width=0.55\textwidth]{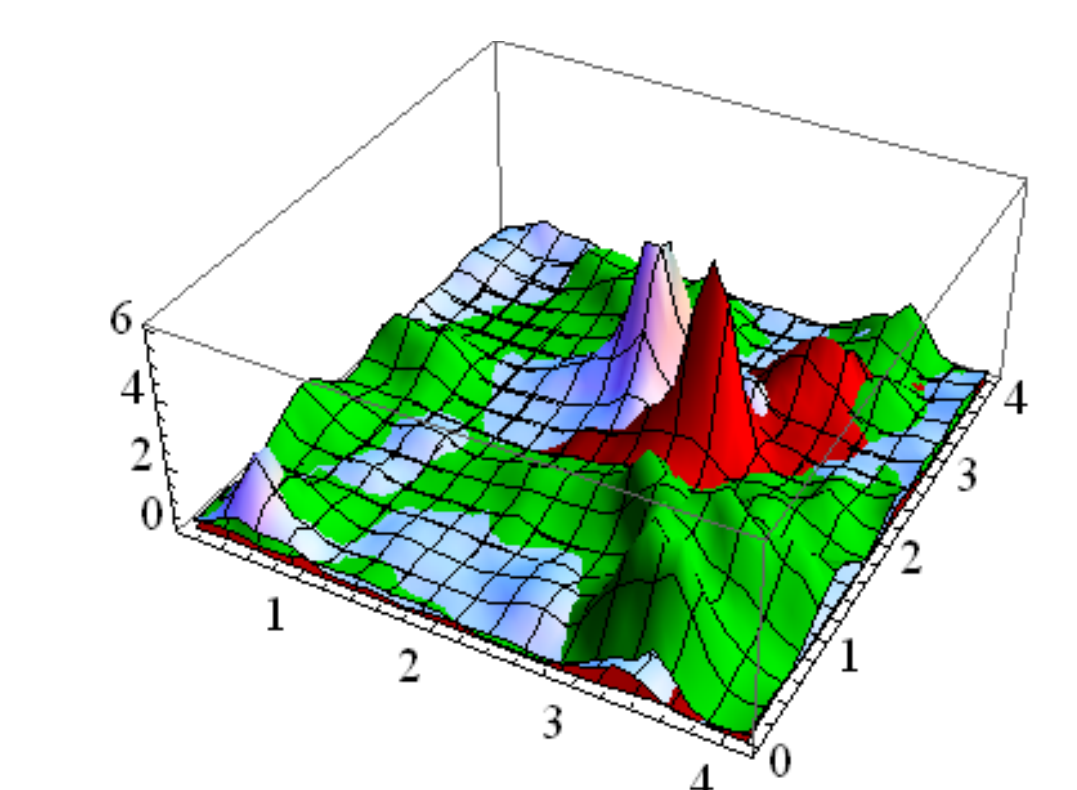}
\caption{ {{\it (Left)}}: dependence of the free energy on Polyakov line values near the deconfinement phase transition, for $SU(3)$ pure gauge theory in which the phase transition is of the first order with a jump, from $0.33=1/3$ at $T<T_c$ to $0.23$ 
at high $T$. 
Parameter $S_0$ on the plot is numerical value of the action per instanton, and 13.25 is its critical value. Figure taken from 
Ref.~\cite{DeMartini:2021dfi}.
{{\it (Right)}}: Space slice of the density distribution of exact zero modes,  from QCD lattice simulaiton at $T\approx T_c$. The three colors
refer to dyons of 
three different types\cite{Larsen:2019sdi}.
}
\label{fig_dyons_from_zero_modes}
\end{center}
\end{figure}

  Details of the solutions themselves, their early applications to QCD and supersymmetric theories can be found in my book \cite{Shuryak:2021vnj}, 
 here  is not a place
to describe those. Perhaps all I can tell here is
that instanton-dyon actions, like that of the instanton
$S_0=8\pi^2/e^2(T)$, decrease as $T\rightarrow T_c$: therefore
they become more numerous, eventually dominating the Polyakov line (free energy of a single quark). This can be seen in the
effective potential for Polyakov line shown in 
left Fig.~\ref{fig_dyons_from_zero_modes} from our work~\cite{DeMartini:2021dfi}, reproducing
the first-order deconfinement transition in quenched (no quarks) $SU(3)$ gauge theory.

 Furthermore, as this approach was extended to 
 QCD with quarks~\cite{Larsen:2015tso,DeMartini:2021xkg} and even versions of ``deformed QCD", 
  the location and strength of $both$ deconfinement and chiral phase transitions were found. And yes, in general deconfinement and chiral transitions are $not$ coincident, and can be moved individually by certain extensions of QCD.

The last thing about instanton-dyons is that one can
 ``go hunting" for  them
in lattice configurations using the {\em fermionic filter method}\footnote{Developed originally by
Gattringer et al and Ilgenfritz et al during 1990's.}  based on {\em zero modes of quark Dirac operator.} Those modes do not see gluons but notice topological solitons.
In Fig.~\ref{fig_dyons_from_zero_modes} (right) we show a sample from Refs.~\cite{Larsen:2018crg,Larsen:2019sdi}.  
Starting with extensive QCD calculations\footnote{Very expensive QCD configurations obtained with top-of the art {\em domain wall fermions} by world-largest supercomputers, yet provided to us for free.}
 %
we search for quark zero modes, locating
  all three types of instanton-dyons  (red,blue and green peaks).

\section{Event-by-event fluctuations and Beam energy scan}
In Ref.~\cite{Shuryak:1997yj}, the event-by-event fluctuations were related to higher thermal susceptibilities, and in 
Ref.~\cite{Stephanov:1998dy},  this method was 
 suggested to be used in a search for hypothetical QCD critical point (CP). In order to do so, one needed {\em beam energy scan} which in due time 
became BES program at RHIC. By now, decades of experimental and theoretical works  seems
to find a location at which certain signals
are present.

 Let us focus here on 
one observable, the ``kurtosis" of nucleon multiplicity distribution. It is 
related to four-nucleon correlations at freeze-out, presumably sensitive to appearance of {\em critical long-range
mode}. Four-nucleon correlations
were studied 
in Ref.~\cite{Shuryak:2018lgd,Shuryak:2019ikv,DeMartini:2020anq} by classical molecular dynamics, novel semiclassical ``flucton" method,
and by direct Path Integral Monte Carlo. The  4-nucleon system is found to be the smallest one, with ``preclusters" decaying into multiple ($\sim 50$) near-zero bound and resonance states. 

Near CP, usual nuclear forces are expected to be appended by  novel diagrams, with two, three and four-body interactions induced by critical mode.  In ref.~\cite{DeMartini:2020anq}, we used lattice results for critical fluctuations near various second-order phases transitions, and analytic results
 by Heidelberg group using
Wetterich's functional RG. 

\begin{figure}[h!]    
        \centering
    \includegraphics[width=0.4\textwidth]{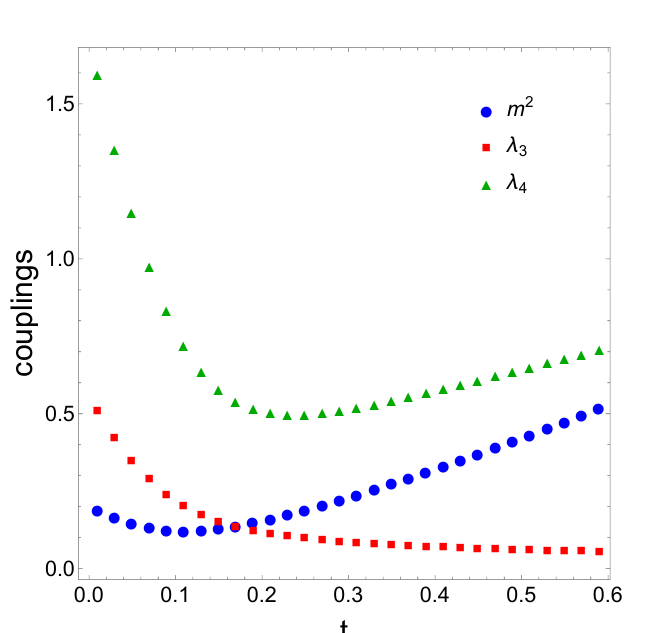}
  \includegraphics[width=0.55\textwidth]{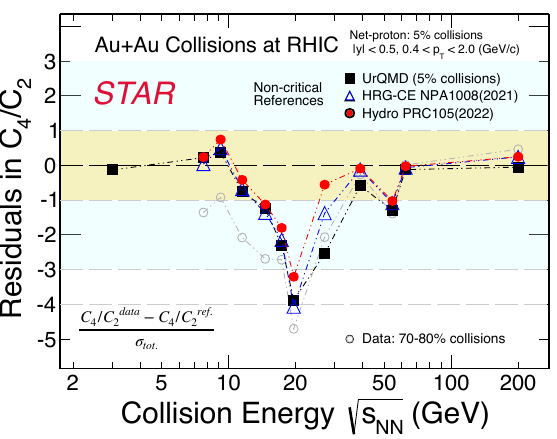}    
    \caption{{{\it (Left)}}: The quadratic, cubic and quartic couplings as a
    function of reduced temperature~\cite{DeMartini:2020anq}. 
{{\it (Right)}}: kurtosis of proton distribution, from STAR collaboration BES-II  run~\cite{STAR:2025zdq}.}
    \label{fig_couplings}
\end{figure}

The derived temperature dependence of 
corresponding couplings is shown
in Fig.~\ref{fig_couplings}~({\it{Left}}), versus the 
reduced temperature 
 $t={T-T_c \over T_c }$.
  Note that
 coefficient of $\phi^2/2$ (squared effective mass $m^2$) is not vanishing at   the CP, $t=0$, although the correlation length is infinite. 
 Triple and quartic coupling, on the other hand, rapidly grow towards the CP ($t\rightarrow 0$).  As a result,  effective free energy $\Delta F$ of a 4-nucleon cluster 
changes sign and $\Delta F$
 gets strongly $repulsive$ in the CP vicinity. 
The resulting predictions of that paper were:
\begin{itemize}
    \item {four-nucleon pre-clustering $\sim exp(-\Delta F/T)$
should be $suppressed$ in the narrow vicinity of the CP.}
\item {This should be observable via multiplicity cumulants, e.g., the kurtosis  $C_4/C_2$ ratio}
\item {Similar effect should also be seen in tritium production, because 4-N clusters decay into shallow $O(50)$
states of $^4He$ which have large branchings into $t+p$,
see \cite{Shuryak:2018lgd} for details.}
\end{itemize}

The latest  set of RHIC BES data 
 \cite{STAR:2025zdq} shown in 
 Fig.~\ref{fig_couplings}~({{\it Right}})
where it also compares with predictions of conventional models (not possessing the critical fluctuations). The deviation between them  reveal
 a narrow dip in kurtosis,  at an energy of $\approx 20$~GeV.

\begin{enumerate}
    \item
The most striking feature is the $sign$ of the 
deviation effect: what is observed is
significant $suppression$ of kurtosis, not an anticipated enhancement in earlier works.
According to  Ref.~\cite{DeMartini:2020anq},
it is due to $repulsive$ manybody forces from diagrams $(b,c)$
predicted to be dominant close to CP.
\item
Furthermore, this dip is rather narrow, corresponding to small reduced temperature $|t|<1/10$. (Note that by
$t$ we mean that in effective Ising model. In heavy ion collisions the Ising plot should be rotated, to tangent to the critical line, so it is in fact combination of the temperature and  chemical potential.)
\item  The dip is located near energy 20~GeV, same location
as suspected previously in another dip in (normalized) tritium production ratio.
\end{enumerate}

\section{Acknowledgments}
 I was a PI and Nuclear Theory group leader from 1993 to 2020: support by DOE Office of Science  over these years is greatly appreciated.
Many friends helped me along the way: I am much indebted to   E.L. Feinberg, I.B. Khriplovich, G.E. Brown, D.I. Diakonov and P. van Baal who are no longer with us. 
Obviously I would reach nowhere without my many collaborators, from Stony Brook
and elsewhere. 

\bibliographystyle{atlasnote}

\bibliography{hi_book4}
\end{document}